\pdfoutput=1
\documentclass[prb,reprint, epsfig,amssymb,bibliography,amsmath,footinbib]{revtex4-2}

\usepackage{blkarray}
\usepackage[dvipsnames]{xcolor}

\usepackage{amssymb,amsmath,bm,epsfig,graphicx,subfigure,hyperref}

\usepackage[utf8]{inputenc}
\usepackage[vietnamese, english]{babel}
\usepackage{chngcntr}
\usepackage{xcolor}
\usepackage{braket}
\usepackage{physics}
\usepackage{sidecap}
\usepackage{lipsum}
\usepackage{amsfonts}
\usepackage{amsmath}
\usepackage{amssymb}
\usepackage{tikz}

\usetikzlibrary{shapes}
\usepackage{xcolor}
\usepackage{wasysym}
\usepackage{array}
\usepackage{bbold}
\usepackage{tabularx}
\usepackage{mathdots}
\usepackage[normalem]{ulem}

\begin{document}
\title{\color{Blue}{Valley-Selective Transport through Structural Domain Walls \\in Rhombohedral Graphene  Quarter Metals}}
\author{\foreignlanguage{vietnamese}{Võ Tiến Phong}$^{1,2,3}$}
\email{vophong@magnet.fsu.edu}
\author{Pablo San-Jose$^{4}$}
\email{pablo.sanjose@csic.es}
\author{Bing Cheng$^{5}$}
\email{bingc287@mit.edu}
\author{Zhengguang Lu$^{1,2,3}$}
\email{zl13b@fsu.edu}
\author{Elsa Prada$^{4}$}
\email{elsa.prada@csic.es}
\author{Long Ju$^{6}$}
\email{longju@mit.edu}
\author{Eugene Mele$^{7}$}
\email{mele@physics.upenn.edu}
\author{Cyprian Lewandowski$^{1,2,3}$}
\email{clewandowski@fsu.edu}
\author{Francisco Guinea$^{8,9}$}
\email{paco.guinea@imdea.org}
\affiliation{$^{1}$Department of Physics, Florida State University, Tallahassee, FL 32306, U.S.A.}
\affiliation{$^{2}$National High Magnetic Field Laboratory, Tallahassee, FL 32310, U.S.A.}
\affiliation{$^{3}$FSU Quantum Initiative, Florida State University, 282 Champions Way, Tallahassee, FL 32306, U.S.A.}
\affiliation{$^{4}$Instituto de Ciencia de Materiales de Madrid (ICMM), CSIC, 28049 Madrid, Spain}
\affiliation{$^{5}$Research Laboratory of Electronics, Massachusetts Institute of Technology, Cambridge, MA 02139, U.S.A.}
\affiliation{$^{6}$Department of Physics, Massachusetts Institute of Technology, Cambridge, MA 02139, U.S.A.}
\affiliation{$^{7}$Department of Physics and Astronomy, University of Pennsylvania, Philadelphia, PA 19104, U.S.A.}
\affiliation{$^{8}$IMDEA Nanoscience, C/ Faraday 9, 28049 Madrid, Spain}
\affiliation{$^{9}$Donostia International Physics Center, Paseo Manuel de Lardiz\'{a}bal 4, 20018 San Sebastián, Spain}
\date{\today}

\begin{abstract}
Domain walls separating quarter metals with opposite valley polarizations have  recently been imaged in rhombohedral graphene multilayers. Depending on their formation mechanism, these walls respond in distinct ways to external magnetic probes. Therefore, understanding their behaviors is critical to the interpretation of experimental results. In this work, we examine domain walls formed from strain-induced one-dimensional structural defects that separate domains with complementary rhombohedral stacking orders. We show that while transport across such domain walls is controlled by the valley degree of freedom, the coupling to a magnetic field is dictated by the momentum-space chirality encoded in the Hamiltonian. As a consequence, structural domain walls are rendered electronically opaque by an applied magnetic field, while intrinsic domain walls formed by spontaneous valley symmetry breaking are unstable in the presence of such an ordering field and can be eliminated. This dichotomy allows the two types of domain walls to be distinguished in a magnetotransport experiment.
\end{abstract}

\maketitle

Rhombohedral graphene multilayers  host weakly dispersive band extrema with Bloch states that are enriched by a significant Berry curvature, rendering them susceptible to a rich assortment of interaction-driven broken symmetry phases. Indeed, many such novel phases have been observed experimentally, including valley-spin ferromagnetism \cite{lee_multicomponent_2016, zhou_half-_2021, arp_intervalley_2024, holleis_fluctuating_2025}, correlated insulators \cite{shi_electronic_2020, hagymasi_observation_2022,liu_spontaneous_2024, han_correlated_2024,han_evidence_2026}, 
anomalous integer and fractional quantum Hall effects \cite{han_large_2024, han_correlated_2024, winterer_ferroelectric_2024, zheng_switchable_2025,  xie_tunable_2025, li_fractional_2025,butler_13_2026},
and superconductivity \cite{zhou_superconductivity_2021, Han2025Chiral, kumar_superconductivity_2025, yang_impact_2025, deng_superconductivity_2025, yang_magnetic_2025, qin_stripe_2025,  nguyen_hierarchy_2025,seo_family_2025,Choi2025-lp, kumar2025pervasive, dutta_reconfigurable_2026, okounkova2026anomalous}. One of the most striking discoveries is the recent observation of robust intrinsic superconductivity from a parent metallic phase with valley-spin polarization \cite{Han2025Chiral, qin_stripe_2025, dutta_reconfigurable_2026}. Theoretical calculations suggest that this state could be a chiral, complex $p$-wave superconductor that hosts topological Majorana zero modes  \cite{Geier2026Chiral, parra-martinez_band_2025, Qin2026Chiral,Yoon2026Quarter}. Partly because chiral superconductivity holds promise for future quantum electronic technologies, significant experimental efforts are underway to further characterize and manipulate this exotic superconducting state.

\begin{figure}
    \centering
    \includegraphics[width=\linewidth]{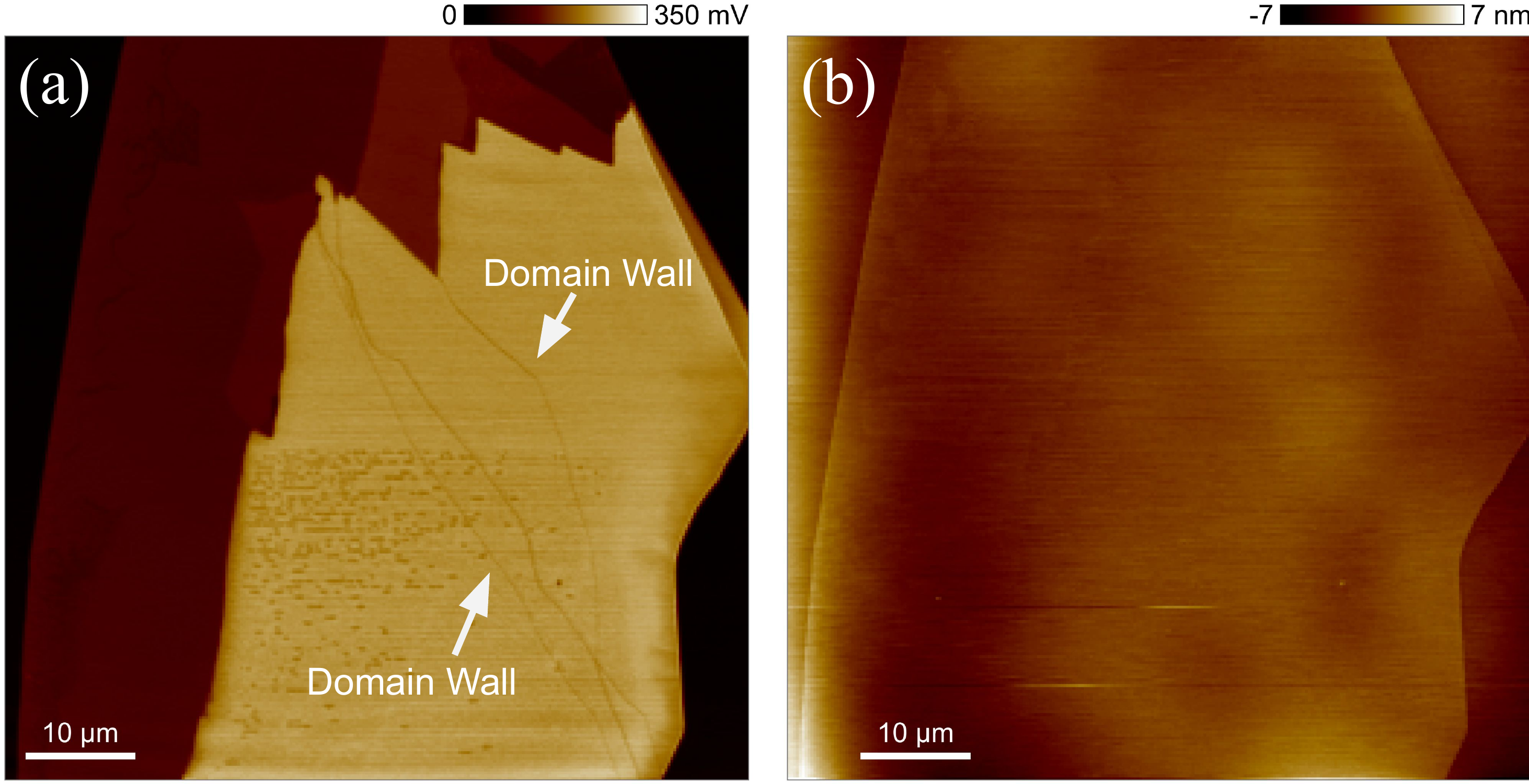}
    \caption{\textbf{Stacking order and topographic characterization of a multilayer flake.} (a) SNOM image of a rhombohedral region (bright yellow) showing the presence of several domain walls. (b) AFM topography of the same region used to identify the number of layers, which is five in this case.  }
    \label{fig:DWimage}
\end{figure}

\begin{figure}
    \centering
    \includegraphics[width=\linewidth]{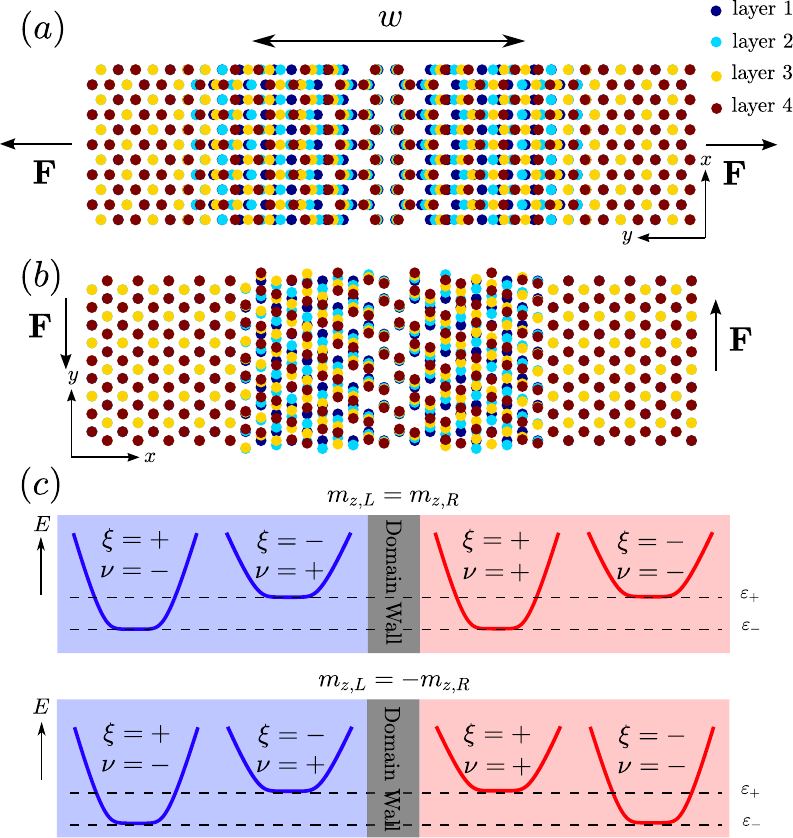}
    \caption{\textbf{Real-space schematic of structural domain walls.} (a) A domain wall of width $w$ running along the zigzag ($x$) direction is created by tensile stress. (b) A domain wall of width $w$ running along the armchair ($y$) direction is created by shear stress. Here, $\mathbf{F}$ is the applied force. (c) Band structure alignment of the two domains separated by a domain wall. Here, $\xi = \pm$ denotes valley while $\nu  =\pm$ denotes chirality. When the $m_z$ mass does not change sign across the domain wall, states in the same valleys spectrally align. On the other hand, when the $m_z$ mass does change sign across the domain wall, states with the same chirality spectrally align. Here, $2|m_z|\approx \varepsilon_+-\varepsilon_-$ is the valley-polarization mass  and is \textit{not} the total gap, which also includes the effect of the displacement field.}
    \label{fig:schematic}
\end{figure}

Recent experiments using magnetic imaging and transport have shown that domain walls separating regions with opposite symmetry-broken order parameters are routinely present in these rhombohedral graphene systems \cite{Han2025Chiral, dutta_reconfigurable_2026, hua2026multi}. Theories predict that these domain walls are phenomenologically much richer than conventional magnetic domain walls due to valley reversal in addition to spin polarization \cite{das2026multicomponent}, and that they may serve as a diagnostic for the nature of the adjacent domains \cite{phong2026valley}. Therefore, elucidating the properties of these domain walls is essential to the understanding and control of the underlying symmetry-broken normal and superconducting phases.

Broadly speaking, we can distinguish between two classes of domain walls \footnote{This is by no means an exhaustive list of domain wall types.}. Line defects that arise from a nonuniform condensation of the order parameter (without poling or pinning by extended real-space defects) are called \textit{intrinsic} domain walls. They are a natural consequence of broken time-reversal symmetry in a magnetic ground state caused by electronic interactions. In principle, these walls can be manipulated as the order parameter is trained by  external fields. In Ref. \cite{phong2026valley}, we  demonstrated that these domain walls between valley-contrasting regions are practically impenetrable to electron transport unless the switching of the valley polarization is accompanied by some form of intervalley mixing near the walls. Similarly, in the superconducting state, these domain walls significantly suppress the Josephson coupling. On the other hand,  \textit{structural} domain walls can be induced by extended line defects in the crystal lattice, most commonly by \textit{lateral} variation of the stacking order across the wall. These do not reflect an underlying spontaneously broken valley symmetry but instead are ``locked in" by the structure of the host lattice. By their nature, these structural domain walls are largely immobile objects \textit{in situ}, although they have been shown to behave as deformable structures under external forces and  fields \cite{Yankowitz:NM14}.  

In this work, we study structural domain walls that are formed by  interior shear and tensile strains. We show that although they exhibit similar transport blockade behavior to intrinsic domain walls, they respond to a training magnetic field in exactly the {\it opposite} way, allowing the two types of domain walls to be distinguished by  magnetotransport measurements. Since this class of structural defects is known to frequently occur in rhombohedral stacks, our work establishes their signatures in transport in the quarter-metal phase.

There are two related quantities that are important to distinguish in our analysis: the valley $\xi=\pm $ and chirality $\nu=\pm $ degrees of freedom. The former labels the projection of the band structure into a definite momentum sector near valley momentum $\mathbf{K}_\xi.$ The variable $\xi$ controls the real-space phase field from the Bloch factor $e^{i\mathbf{K}_\xi \cdot \mathbf{r}}.$ The chirality refers to the phase winding of the off-diagonal elements of the valley-projected Hamiltonian $k_x+ i\nu k_y$. This also corresponds to the pseudospin winding around the valley. The variable $\nu$ thereby dictates the relative signs of the orbital magnetization and the Berry curvature, and hence determines the coupling of the electronic states to an external magnetic field \cite{PhysRevLett.95.137205,PhysRevLett.95.137204, PhysRevB.74.024408,PhysRevLett.99.197202}. Note that for an intrinsic domain wall, $\xi = \nu$ on both sides because the crystal stacking order does not change, so this distinction is unnecessary, as in Ref. \cite{phong2026valley}. However, with a \textit{structural} domain wall, this is now a meaningful and necessary distinction, since the valley-chirality relation depends on the stacking. A key conclusion in this work is that the transparency of a wall is controlled by the Bloch phases parameterized by $\xi$, whereas an external magnetic probe couples to $\nu$.

We begin with measurements that demonstrate the existence of structural domain walls in mechanically exfoliated multilayer graphene. Fig. \ref{fig:DWimage}(a) shows the third-harmonic near-field image measured by scanning near-field optical microscopy (SNOM) on a five-layer graphene flake, while the corresponding atomic force microscopy (AFM) topography is shown in Fig. \ref{fig:DWimage}(b). The AFM image is relatively uniform across the flake, consistent with a homogeneous five-layer thickness. In contrast, the SNOM image reveals pronounced spatial contrast associated with different stacking configurations. The bright-yellow region corresponds to rhombohedral stacking, within which several extended line-like features are clearly visible. These features are structural domain walls that separate adjacent regions with possibly complementary rhombohedral stacking orders. Such domain walls are commonly observed in mechanically exfoliated rhombohedral multilayer graphene, indicating that structural defects are routinely present in these samples.

Inspired by these observations, we now turn to the theoretical description of these domain walls and their transport signatures. We consider structural domain walls along two high-symmetry crystallographic directions: one that extends along the zigzag direction (along the $x$ axis) produced by tensile strain, as shown in Fig. \ref{fig:schematic}(a), and one that extends along the armchair direction (along the $y$ axis) from shear strain, as shown in Fig. \ref{fig:schematic}(b). In both cases, the domain wall smoothly interpolates between a $CBA\dots$ stacking sequence on one side and an $ABC\dots$ stacking sequence on the other. In Supplementary Material (SM) Sec. \ref{sec: Continuum elasticity modeling of quenched domain walls}, we consider general domain wall configurations using continuum elasticity theory and varying the number of coupled layers. According to our theory, there exist several possible structures for these domain walls, such as $AA$ or saddle point (SP) boundaries of a tensile, shear, or intermediate kind. They have a typical width of  $w\sim 10$ nm. In the main text, we focus on a case where the domain wall runs through an $AA$ region, whose construction is described in SM Sec. \ref{sec: Lattice Structures}.


To analyze transport, we first use an analytical continuum model assuming sharp walls, and then validate the results with more detailed numerical simulations \cite{PhysRevB.88.115409, san2014stacking, PhysRevB.91.214113}. The surface-projected continuum Hamiltonians on the two sides of  the domain wall at valleys  $\mathbf{K}_\xi = \xi 4\pi/3a \hat{x}$ are given by \footnote{Here, ``left'' means either $x<0$ or $y <0$ and ``right'' means either $x>0$ or $y>0.$ In the cases where $y$ is the direction orthogonal to the domain wall, it may be more appropriate to use ``below'' and ``above''. Also, we shall always order the basis in such a way that the layers are fixed} 
\begin{equation}
\label{eq: real space Hamiltonian}
\begin{split}
    \mathcal{H}_{s,\xi}(\mathbf{k}) &= \begin{pmatrix}
        \delta_{s,\xi} & \alpha \left(\xi k_x-i s k_y  \right)^N \\
        \alpha \left(\xi k_x+ i s k_y  \right)^N & -\delta_{s,\xi}
    \end{pmatrix},
\end{split}
\end{equation}
where $\alpha = \hbar^N v_F^N/\gamma_1^{N-1},$ $v_F$ is the Fermi velocity of graphene,  $\gamma_1$ is the direct hopping between eclipsed sites, $\xi = \pm$ denotes the valleys, and $s=-$ (or $s = L$) denotes the left side and $s=+$ (or $s = R$) denotes the right side of the domain wall. As a consequence of the change in stacking order across the domain wall, on the left side, $\nu = \mp$ is locked to $\xi = \pm,$ while on the right $\nu = \pm$ is locked to $\xi = \pm$; in brief, $\nu = s\xi$. This is shown spectrally in Fig. \ref{fig:schematic}(c) in the presence of a valley polarization. Here, $\delta_{L/R,\xi}$ are gaps that incorporate the effects of both the displacement field and spontaneous valley symmetry breaking. Henceforth, we set $\alpha = 1$ and measure momentum and energy in dimensionless units. It is important to note that the basis for $\mathcal{H}_{L,\xi}$ is localized on the $(B_1, A_N)$ sublattices while the basis for $\mathcal{H}_{R,\xi}$ is localized on the $(A_1,B_N)$ sublattices.  This is because, in the absence of the diagonal terms, the two sides of the domain wall are related to each other by a $C_{2z}$ rotation: for $\delta_{L,\xi} = \delta_{R,\xi} = 0$, $\mathcal{H}_{L,\xi}(\mathbf{k}) = \mathcal{H}_{R,-\xi}(-\mathbf{k}).$

In the continuum model, valley polarization is modeled by including a valley-dependent mass in addition to the valley-independent gap induced by the displacement field. To model the \textit{same} valley polarization on the two sides of a domain wall, we let the gaps on the two sides in the \textit{same} valleys be identical: $\delta_{L,\xi} = \delta_{R,\xi}.$ On the other hand, to model the \textit{different} valley polarizations on the two sides of a domain wall, we let the gaps on the two sides in different valleys be identical: $\delta_{L,\xi} = \delta_{R,-\xi}.$  These observations lead to an important consequence: tunneling between the two sides with opposite chiralities is \textit{not} kinematically prohibited even in the zigzag direction.  This is diametrically opposed to the blockade behavior studied in Ref. \cite{phong2026valley} where valley and chirality are locked. The origin of these contrasting behaviors is, of course, the relative orientation of the two domains under consideration.

\begin{figure}
    \centering
    \includegraphics[width=1\linewidth]{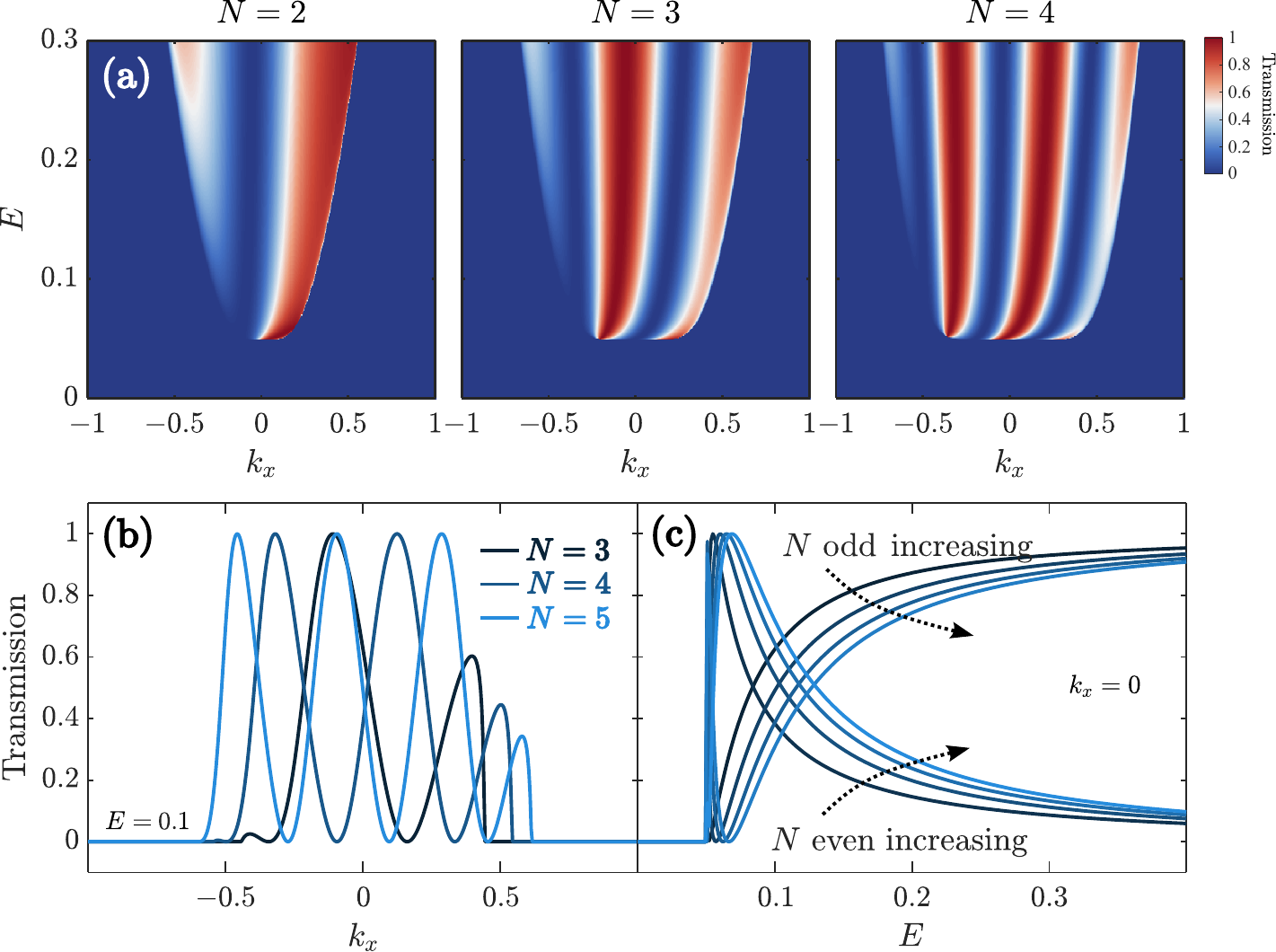}
    \caption{\textbf{Transmission through an abrupt domain wall in the simplified two-band continuum model.} (a) Transmission as a function of energy $E$ and parallel momentum $k_x$ for different layer numbers. (b) Energy cross section at $E=0.1$ as a function of $k_x$ for different layer numbers. (c) Transmission for normal incidence ($k_x = 0$) as a function of energy for different layer numbers. At larger energies in the window shown, the transmission for odd $N$ approaches one, while the transmission for even $N$ approaches zero.  In all cases, the transmission is set to zero outside of the bulk spectrum. Here, $\xi = +$ on both sides of the wall, $\delta = 0.05$, and energies and momenta are measured in units of $\gamma_1 = 380$ meV and $\gamma_1/\hbar v_F = 0.142/a$ (with $a$ the lattice constant), respectively.}
    \label{fig:fig1}
\end{figure}

 Concretely, we now show that the blockade behavior is controlled by $\xi$ and not $\nu$.  We can examine transmission analytically in the zigzag direction. In this case, the two valleys are projected to different momentum sectors that are separated from each other by $4\pi/3a$, where $a$ is the lattice constant. We assume that $\delta_{L,+}=\delta_{R,+} = \delta < \delta_{L,-}=\delta_{R,-},$ so we focus on only the $\xi = +$ valley by assuming the chemical potential is below the band minimum of the minority $\xi = -$ valley. The two domains have \textit{opposite} chiralities in the same valley.  The domain wall is located at $y = 0.$ The bulk eigenfunctions for Hamiltonian \eqref{eq: real space Hamiltonian} can be found piecewise for negative and positive $y$, respectively given by
\begin{equation}
\begin{split}
    \ket{\phi_L} &=  e^{ik_yy}\begin{pmatrix}
        \delta+E \\
        k_-^N
    \end{pmatrix}, \quad  
    \ket{\phi_R}  = e^{ik_yy}\begin{pmatrix}
        \delta+E \\
        k_+^N
    \end{pmatrix},   
\end{split}
\end{equation}
where $E = \pm\sqrt{(k_y^2+k_x^2)^N+\delta^2}$ and $k_\pm = k_x \pm i k_y.$ To proceed, we need to implement the proper matching condition exactly at $y=0.$ This is most transparently done in the microscopic layer-sublattice basis wherein the matching condition is simply continuity of the wavefunction. To build the scattering states, we note that at a fixed $E$ and $k_x,$ there are generically $2N$ $k_y$ wavevectors, two of which are real, denoted $\pm k_y$, and the rest are complex, denoted $\kappa_{y,n\in\lbrace 1,...,N-1\rbrace}$ and $\kappa^*_{y,n\in\lbrace 1,...,N-1\rbrace}$. On the left side, the scattering state is built out of an incoming $k_y$ wave, a reflected $-k_y$ wave, and $N-1$ normalizable evanescent waves, written as linear combinations of the basis states in Eq. \eqref{eq: basis 1} in SM Sec. \ref{sec: Scattering States}. On the right side, the scattering state is built out of a transmitted $k_y$ wave and $N-1$ normalizable evanescent waves, written as linear combinations of the basis states in Eq. \eqref{eq: basis 2} in SM Sec. \ref{sec: Scattering States}. Since the basis states in Eqs. \eqref{eq: basis 1} and \eqref{eq: basis 2} are written in a common real-space representation, we can match the scattering states on left and right sides by simply setting them equal exactly at $y = 0.$ This allows us to solve for the transmission across the domain wall.

\begin{figure}
    \centering
    \includegraphics[width=\linewidth]{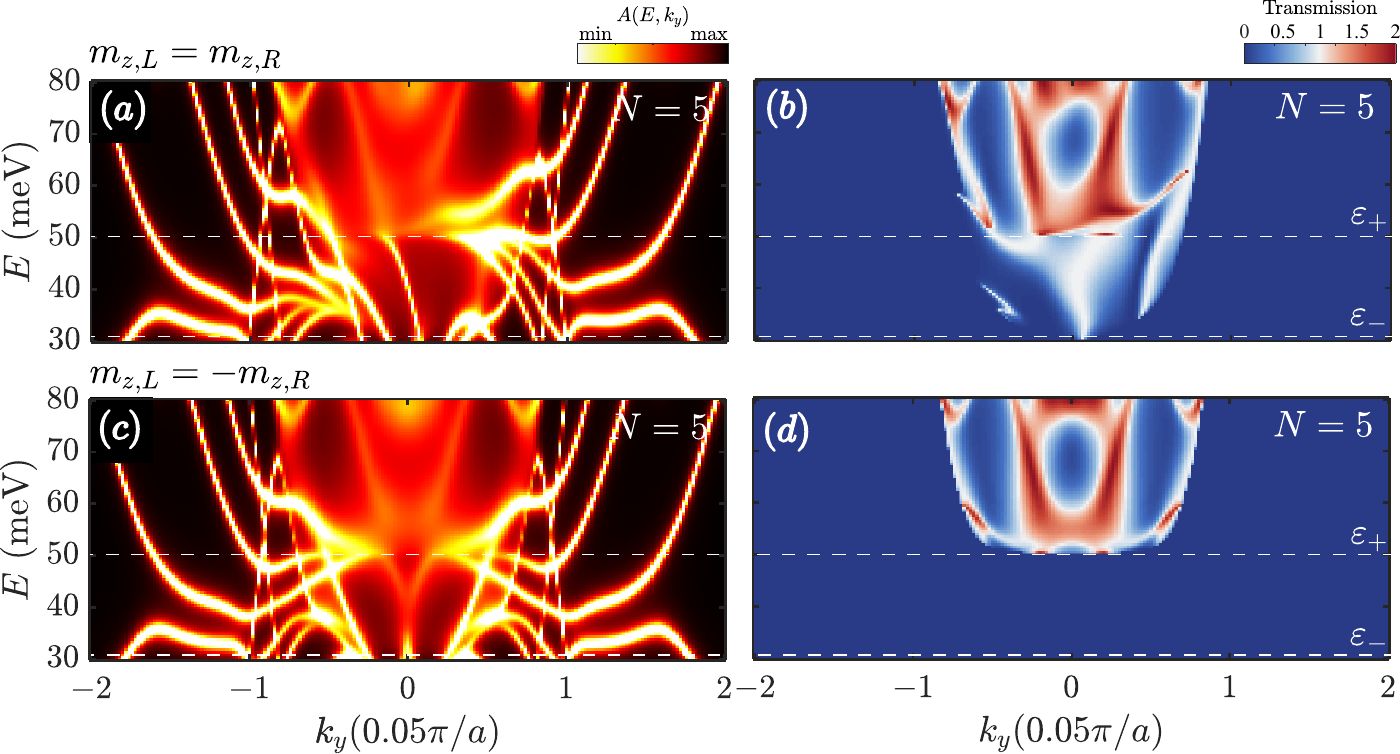}
    \caption{\textbf{Transmission through a finite-width domain wall along the armchair direction for $N=5$}. (a) Spectral function $A(E,k_y)$ and (b) transmission  across the armchair domain wall where $m_{z,L} = m_{z,R}$. (c) Spectral function $A(E,k_y)$ and (d) transmission  across an armchair domain wall where  $m_{z,L} = -m_{z,R}$. Here, $m_z = 10$ meV, $\Delta = 20$ meV, and $w = 50a \approx 12$ nm.}
    \label{fig:fig4}
\end{figure}

The results for the transmission through an abrupt structural domain wall calculated in this simplified single-channel continuum model are shown in Fig. \ref{fig:fig1}. In Fig. \ref{fig:fig1}(a), we show the transmission as a function of energy $E$ and parallel momentum $k_x$ for different layer numbers $N = 2,3,4.$ At a fixed energy within the bulk-state region, we observe  oscillations in the transmission as $k_x$ is varied, as shown in Fig. \ref{fig:fig1}(b). The frequency of oscillations increases monotonically as a function of $N.$ Numerically, we observe that the transmission reaches unity $N-2$ times within the bulk-state region shown in Fig. \ref{fig:fig1}(a). We note the clear lack of $k_x \mapsto -k_x$ symmetry in the transmission function due to valley polarization. In Fig. \ref{fig:fig1}(c), we show the transmission as a function of energy for normal incidence $k_x = 0.$ We observe two distinct behaviors based on the parity of  $N.$ For $N$ odd, the transmission is minimal near the band edge and approaches unity at larger energies in the energy range shown. For $N$ even, the trend is reversed. These results illustrate that a structural domain wall can mediate significant electron transport even when the two domains carry opposite chiralities, illustrating that the blockade behavior is not controlled by chirality $\nu.$

The odd-even $N$ behavior originates from the simultaneous reversal of momentum-space chirality and exchange of the low-energy microscopic basis across the structural domain wall. At normal incidence in the $E\rightarrow \infty$ limit, the evanescent modes do not play a crucial role; only the propagating modes are active. Reversing the chirality interchanges $k_-^N$ and $k_+^N$, producing the relative factor $k_+^N/k_-^N = (-1)^N.$ The accompanying exchange of the active surface-sublattice basis, $(B_1,A_N)\leftrightarrow (A_1, B_N)$, contributes an additional minus sign. The total resulting relative phase is therefore $(-1)^{N+1}$. For odd $N$, this factor is $+1$, allowing constructive matching of the microscopic wavefunctions and favoring transmission. For even $N$, it is $-1$, producing destructive interference and suppressing normal-incidence transmission. This behavior is reminiscent of the property at the interface between two topological insulators with reversed Dirac velocities \cite{Takahashi2011Gapless}. For $E \gtrsim \delta,$ the dependence on $N$ is flipped because, in this regime, transmission is determined by the interference of the $N-1$ evanescent modes on each side of the wall.

We now generalize our results to more realistic settings using an atomistic tight-binding model. We use the real-space Slater-Koster parametrization of the carbon-carbon hopping amplitude from Ref. \cite{Moon2013Optical}. With this, we can model the domain walls by modulating the equilibrium positions of the carbon atoms within a width $w$ subjected to shear or tensile stress as specified earlier and shown in Fig. \ref{fig:schematic}.  Valley polarization is implemented by a Haldane-like interaction that has the same sign on both sublattices, as done in Ref. \cite{phong2026valley}. This interaction is imposed only on layer 1, $\delta\mathcal{H}_{\tau_z}(\mathbf{k}) = \frac{2m_z}{3\sqrt{3}} \sum_{j=1}^3 \sin \left[\mathbf{k} \cdot \mathbf{a}_j \right],$ where $\mathbf{a}_{j+1} = a\left(\cos\left[2\pi j/3 \right],\sin\left[2\pi j/3 \right] \right)$. This choice is  convenient because layer 1 is unstrained in our construction, so we do not need to consider the spatial modulation of the Haldane-like term. Finally, we add a layer-progressive potential energy $\Delta$ induced by the perpendicular displacement field, with the sign chosen so that conduction electrons preferentially populate layer 1. The tight-binding approach allows many scattering channels at a fixed energy and can, in principle, handle a domain wall along any crystallographic axis; for simplicity, we consider only the armchair and zigzag directions. Furthermore, we allow the possibility of $m_z$ changing sign across the domain wall, $m_{z,L} = \pm m_{z,R}.$ In the main text, we focus on the armchair direction since the zigzag direction has already been considered in the simplified continuum model. The tight-binding results for the zigzag direction are shown in SM Sec. \ref{sec: Atomistic Tight-Binding Model} and are consistent with the continuum-model results.

For an armchair domain wall of width $w = 50a \approx 12$ nm, the results are shown in Fig. \ref{fig:fig4} for $N=5$ where $m_z$ has the same (a,b) and different  signs (c,d) across the walls. Similar results for $N = 3,4$ are shown in  SM Sec. \ref{sec: Atomistic Tight-Binding Model}. Along the armchair direction, both valleys are projected to the same momentum sector around $k_y = 0.$ In Fig. \ref{fig:fig4}(a,c), we show the spectral function of the domain walls, where in addition to the bulk continuum states, we also observe the proliferation of many dispersive wall-localized states induced by the modification of domain-wall bonds due to shear strain \cite{zhang2013valley,ju2015topological}. We observe two conduction band minima at $\varepsilon_\pm$  because of valley polarization. For $m_{z,L} = m_{z,R},$ the same valleys on the two sides of the domain wall are spectrally aligned but the chiralities are mismatched. We find two regimes: for $\varepsilon_- \leq E < \varepsilon_+,$ the transmission comes entirely  from states at the $\xi=+$ valley, and for $E \geq \varepsilon_+,$ there is transmission from both valleys. This validates again that a mismatch in chirality does \textit{not} dictate blockade behavior. For $m_{z,L} = -m_{z,R},$ opposite valleys on the two sides of the domain wall are now spectrally aligned. In the energy window $\varepsilon_- \leq E<\varepsilon_+$, states on both sides of the domain wall have the same chirality. Despite that, we observe from Fig. \ref{fig:fig4}(d) that there is practically zero transmission for $E < \varepsilon_+.$ Even though  transmission is not kinematically prohibited since both valleys are mapped to the same momentum, the near-zero value here clearly demonstrates that the blockade behavior is controlled by spectrally aligning states with the same $\xi$; chirality $\nu$ is not the determining factor here. In other words, the suppression of transmission is due to destructive interference of the wavefunction in real space originating from the phase factors $e^{i \mathbf{K}_\xi \cdot \mathbf{r}}.$

The above considerations immediately suggest an experimental magnetic probe to distinguish the two types of domain walls. Such a probe operates on the dichotomous principle that transport is controlled by valley $\xi$ but magnetic coupling is controlled by chirality $\nu$ \cite{PhysRevLett.95.137205,PhysRevLett.95.137204, PhysRevB.74.024408,PhysRevLett.99.197202}. On one hand, an intrinsic domain wall is defined by having opposite valley polarizations on the two sides, which corresponds to having opposite chiralities and therefore exhibits suppressed transmission (practically zero in the absence of intervalley mixing). An external magnetic field couples to chirality; thus, above the appropriate switching field, and provided that the intrinsic wall is not strongly pinned, the field-misaligned domain is expected to reverse, restoring transmission to the case of a homogeneous sample. On the other hand, a structural domain wall may have opposite or equal valley polarization on the two sides, corresponding to equal or opposite chiralities, respectively. An external field energetically favors the establishment of uniform chirality across the boundary. This configuration corresponds to having opposite valley polarizations on the two sides; a structural wall that is transparent at zero field therefore becomes opaque as the field increases, and one that is opaque at zero field remains opaque at finite fields. In brief, above a critical field strength, an intrinsic domain wall would be eliminated and conduction would consequently be enhanced, while a structural domain wall would become electronically opaque.

In conclusion, our work demonstrates that different types of domain walls in symmetry-broken rhombohedral systems respond in distinct manners to external probes. Understanding these characteristic behaviors not only reveals the nature of the domains separated by such walls, but might also offer important clues to their future design and manipulation for integrating symmetry-broken phases into quantum electronic circuits \cite{gubbiotti20252025}.

 V.T.P. and C.L. were supported in part by start-up funds from Florida State University and the National High Magnetic Field Laboratory and in part by NSF CAREER grant No. DMR-2543710. E.P. and P.S-J. acknowledge support by the Spanish AEI through Grant No. PID2024-161665NB-I00 and Severo Ochoa Centers of Excellence program (CEX2024-001445-S), funded by MICIU/AEI/10.13039/501100011033, ``ERDF A way of making Europe'' and ``ESF+''. B.C. and L.J. acknowledge funding from the US Department of Energy, Office of Science, Basic Energy Sciences, Materials Sciences and Engineering Division under grant no. DE-SC0025325. 
 IMDEA Nanociencia acknowledges support from the Spanish State Research Agency (AEI/10.13039/501100011033) and the Ministry of Science, Innovation and Universities (MICIU) through the ``Severo Ochoa" Programme for Centres of Excellence in R\&D (CEX2020-001039-S and CEX2025-001660-S). 
F. G. acknowledges support from NOVMOMAT, project PID2022-142162NB-I00 funded by MICIU/AEI/10.13039/501100011033 and by FEDER, UE.
 F.G. acknowledges funding from the EU NextGenerationEU/PRTR-C17.I1, the Basque Government’s Department of Education through project PIBA\_2023\_1\_0007 (STRAINER), as well as by the IKUR Strategy under the collaboration agreement between Ikerbasque Foundation and DIPC on behalf of the Department of Education of the Basque Government. Work by E.J.M. is supported by the Department of Energy grant DE-FG02-84ER45118. Numerical calculations are done using the High Performance Compute Cluster  of the Research Computing Center (RCC) at Florida State University.


\bibliography{references,domain_wall_refs}

\clearpage

\setcounter{equation}{0}
\setcounter{figure}{0}
\renewcommand{\theequation}{S\arabic{equation}}
\renewcommand{\thefigure}{S\arabic{figure}}
\renewcommand{\thetable}{S\Roman{table}}
\renewcommand{\bibnumfmt}[1]{[#1]}
\renewcommand{\citenumfont}[1]{#1}

\onecolumngrid

\onecolumngrid
\begin{center}
\Large
    Supplementary Material
\end{center}


 \section{Lattice Structures}
 \label{sec: Lattice Structures}

For Fig. \ref{fig:schematic} in the main text, a  rhombohedral $N$-layer graphene stack is constructed by shifting layer $\ell \in \left[1,2,...,N\right]$ by $\pm (\ell-1)a/\sqrt{3} \hat{y},$ where $a$ is the lattice constant of graphene. In our convention, the zigzag (armchair) direction runs along the $x$  $(y)$ axis. To construct a domain wall, we  modulate the shift registry from one side of the domain wall to the other, as shown in Fig. \ref{fig:schematic}(a,b). For a domain wall running along the zigzag direction, we modulate the shift registry by $g(y) (\ell-1)a/\sqrt{3} \hat{y},$ with $g(y)\approx\tanh(y/w),$ where $w$ is the width of the domain wall. The bonds inside the domain wall experience \textit{tensile} stress. For a domain wall running along the armchair direction, we modulate the shift registry by $g(x) (\ell-1)a/\sqrt{3} \hat{y}.$ The bonds inside the domain wall experience \textit{shear} stress. As a very rough estimate, the fractional strain is given by $4a/\sqrt{3}w,$ which is about $6\%$ for a $w\approx10$ nm domain wall. In our construction, chosen for the convenience of numerical simulation, the $\ell = 1$ layer is not strained at all while the $\ell = N$ layer is maximally strained. Alternative constructions can include situations where all layers are strained in different ways or one of the inner layers is unstrained. However, we believe that none of our principal conclusions depends on the precise nature of strain since they only concern the symmetries of the wavefunctions from the bulk domains.

\section{Scattering states}
\label{sec: Scattering States}

We measure crystal momentum in units of $\gamma_1/\hbar v_F = 0.142/a$ and energy in units of $\gamma_1 = 380$ meV. The bulk basis states for Hamiltonian \eqref{eq: real space Hamiltonian} for the left side are:
\begin{equation}
\label{eq: basis 1}
\begin{split}
    \ket{\psi_{B_1}(\mathbf{k})} &= \frac{(-1)^{N+1}}{\mathcal{A}(\mathbf{k)}}\begin{pmatrix}
        1 &-   k_-  &\left[-   k_- \right]^2 & \dots
    \end{pmatrix}^T_\ell \otimes \begin{pmatrix}
        0 &
        1
    \end{pmatrix}_\sigma^T ,  \\
    \ket{\psi_{A_N}(\mathbf{k})} &= \frac{1}{\mathcal{A}(\mathbf{k)}}\begin{pmatrix}
        \dots & \left[-  k_+\right]^2 & -  k_+ & 1 
    \end{pmatrix}^T_\ell \otimes \begin{pmatrix}
        1 &
        0
    \end{pmatrix}_\sigma^T.
\end{split}
\end{equation}
The basis states for the right side are:
\begin{equation}
\label{eq: basis 2}
\begin{split}
    \ket{\psi_{A_1}(\mathbf{k})} &= \frac{(-1)^{N+1}}{\mathcal{A}(\mathbf{k})}\begin{pmatrix}
        1 & -   k_+ & \left[-  k_+ \right]^2 & \dots
    \end{pmatrix}^T_\ell\otimes \begin{pmatrix}
        1 &        0
    \end{pmatrix}_\sigma^T, \\
    \ket{\psi_{B_N}(\mathbf{k})} &= \frac{1}{\mathcal{A}(\mathbf{k)}}\begin{pmatrix}
    \dots & \left[-  k_-\right]^2& -  k_- & 1 
    \end{pmatrix}^T_\ell\otimes \begin{pmatrix}
        0 &        1
    \end{pmatrix}_\sigma^T.    
\end{split}
\end{equation}
Here, $\ell$ denotes layer, $\sigma$ denotes in-plane sublattice, and $\mathcal{A}(\mathbf{k})$ is the normalization factor. Assuming that we are within an energy window with only one scattering channel on each side, the full wavefunction on the left side can be written as
\begin{equation}
\begin{split}
    \ket{\Psi_{E,k_x}(y<0)} &= e^{ik_yy} \left[ (\delta+E) \ket{\psi_{B_1}(k_x,k_y)} + (k_x-ik_y)^N\ket{\psi_{A_N}(k_x,k_y)} \right] \\
    &+ \mathfrak{r}e^{-ik_yy} \left[ (\delta+E) \ket{\psi_{B_1}(k_x,-k_y)} + (k_x+ik_y)^N\ket{\psi_{A_N}(k_x,-k_y)} \right]  \\
    &+\sum_{n=1}^{N-1} \mathfrak{a}_ne^{i\kappa_{y,n}y} \left[ (\delta+E) \ket{\psi_{B_1}(k_x,\kappa_{y,n})} + (k_x-i\kappa_{y,n})^N\ket{\psi_{A_N}(k_x,\kappa_{y,n})} \right].
\end{split}
\end{equation}
The wavefunction on the right side can be written similarly as
\begin{equation}
\begin{split}
    \ket{\Psi_{E,k_x}(y>0)} &=  \mathfrak{t}e^{ik_yy} \left[ (\delta+E) \ket{\psi_{A_1}(k_x,k_y)} + (k_x+ik_y)^N\ket{\psi_{B_N}(k_x,k_y)} \right]  \\
    &+\sum_{n=1}^{N-1} \mathfrak{b}_ne^{i\kappa_{y,n}^*y} \left[ (\delta+E) \ket{\psi_{A_1}(k_x,\kappa^*_{y,n})} + (k_x+i\kappa^*_{y,n})^N\ket{\psi_{B_N}(k_x,\kappa^*_{y,n})} \right].
\end{split}
\end{equation}
Here, the complex wavevectors $\kappa_{y,n}$ are chosen to have negative imaginary parts for normalizability. By setting $    \lim_{y\rightarrow 0} \ket{\Psi_{E,k_x}(y<0)} = \lim_{y \rightarrow 0} \ket{\Psi_{E,k_x}(y>0)},$
we can solve for $\lbrace \mathfrak{r},\mathfrak{a}_n, \mathfrak{t}, \mathfrak{b}_n \rbrace.$ The reflection and transmission are given by $\mathfrak{R} = |\mathfrak{r}|^2$ and $\mathfrak{T} = |\mathfrak{t}|^2$, respectively. It is worth emphasizing that the existence of at most one propagating wave in either direction at any energy $E$ is particular to the simplified model we are considering. In a more realistic model, there are generically multiple degenerate scattering channels. Nonetheless, our simplified single-channel model is enough to illustrate the main physics.



\section{Atomistic tight-binding model}
\label{sec: Atomistic Tight-Binding Model}

In this section, we describe the lattice structures that feature structural domain walls, either along the zigzag or armchair direction. A pristine $AA$ multilayer graphene lattice is obtained by a regular array of lattice points on two sublattices situated at the following locations
\begin{equation}
\begin{split}
    \mathbf{r}_{A,\ell} &= n_1 \mathbf{a}_1 + n_2 \mathbf{a}_2 + \left(\ell-1 \right)\mathbf{d}, \\
    \mathbf{r}_{B,\ell} &= n_1 \mathbf{a}_1 + n_2 \mathbf{a}_2 + \mathbf{b}_0 + \left(\ell-1 \right)\mathbf{d},
\end{split}
\end{equation}
where $\mathbf{a}_1 = a \left(1,0,0\right) , \mathbf{a}_2 = a \left(1/2,\sqrt{3}/2,0\right) , \mathbf{d} = \left(0,0,d \right), \mathbf{b}_0 = \left(0,a/\sqrt{3},0 \right),$ $a=2.46$ \AA, and $d=3.34$ \AA. Here, $\ell \in [1,2,...,N]$ is the layer index. The tight-binding Hamiltonian is represented by
\begin{equation}
\begin{split}
     \hat{\mathcal{H}}_0 &= \sum_{\mathbf{r}\neq\mathbf{r}'} t\left(\mathbf{r}- \mathbf{r}'\right) \hat{c}^\dagger(\mathbf{r})\hat{c}(\mathbf{r}') \\
     &+ \sum_{\mathbf{r}} \left[ \frac{\Delta(N-1)}{2} - \Delta (\ell-1) \right] \hat{c}^\dagger(\mathbf{r})\hat{c}(\mathbf{r}),   
\end{split}
\end{equation}
where the sum is over all lattice sites from both sublattices and layers, $\hat{c}^\dagger(\mathbf{r})$ is a fermionic creation operator at site $\mathbf{r},$ $\Delta$ is the on-site energy  induced by the displacement field, and \cite{Moon2013Optical} 
\begin{equation}
\begin{split}
    t(\mathbf{r}) = &-\left(2700 \text{ meV} \right) \exp \left( - \frac{r-a/\sqrt{3}}{0.184a} \right) \left[1 -  \frac{r_z^2}{r^2} \right]\\
    &+\left(480 \text{ meV} \right) \exp \left( - \frac{r-d}{0.184a} \right) \left[  \frac{r_z^2}{r^2} \right],
\end{split}
\end{equation}
where $r = |\mathbf{r}|$ and $r_z = \mathbf{r} \cdot \hat{z}.$ Now, we implement valley polarization using a Haldane-like interaction
\begin{equation}
    \delta\hat{\mathcal{H}}_{\tau_z} = \frac{m_z}{3\sqrt{3}i}\sum_{\mathbf{r},\ell=1,\mathbf{r}_{i=1,2,...,6}} (-1)^{\theta_i}\hat{c}^\dagger(\mathbf{r}) \hat{c}(\mathbf{r} + \mathbf{r}_i),
\end{equation}
where $\theta_i = 0$ for $i$ odd, $\theta_i = 1$ for $i$ even, and $\mathbf{r}_{j+1} = a \left( \cos \left[\pi j/3\right],\sin \left[ \pi j/3 \right] \right).$ We only allow hoppings within the same layer or between adjacent layers; in all cases, the range is less than 4$a$. Strain is implemented in the following way: $\mathbf{r} \mapsto \mathbf{r}+ \delta \mathbf{r}(\mathbf{r} = (x,y)),$ where
\begin{equation}
    \begin{split}
        \delta \mathbf{r}(x,y) &= g(x,y) (\ell-1) \mathbf{b}_0, \\
        g(x,y) &\approx  \left\{\begin{array}{ll}
      \tanh \left((x-x_0)/w\right) & \text{if armchair domain wall} \\
      \tanh \left((y-y_0)/w\right) & \text{if zigzag domain wall} \\
\end{array} \right.,
    \end{split}
\end{equation}
$w$ is the width of the domain wall and $(x_0,y_0)$ is the center of strain, chosen to make the sample as symmetric as possible. In numerical calculation, we use a linearized version of the hyperbolic tangent. We note that $\delta \mathbf{r} = \mathbf{0}$ for $\ell = 1.$ 

\begin{figure*}
    \centering
    \includegraphics[width=\linewidth]{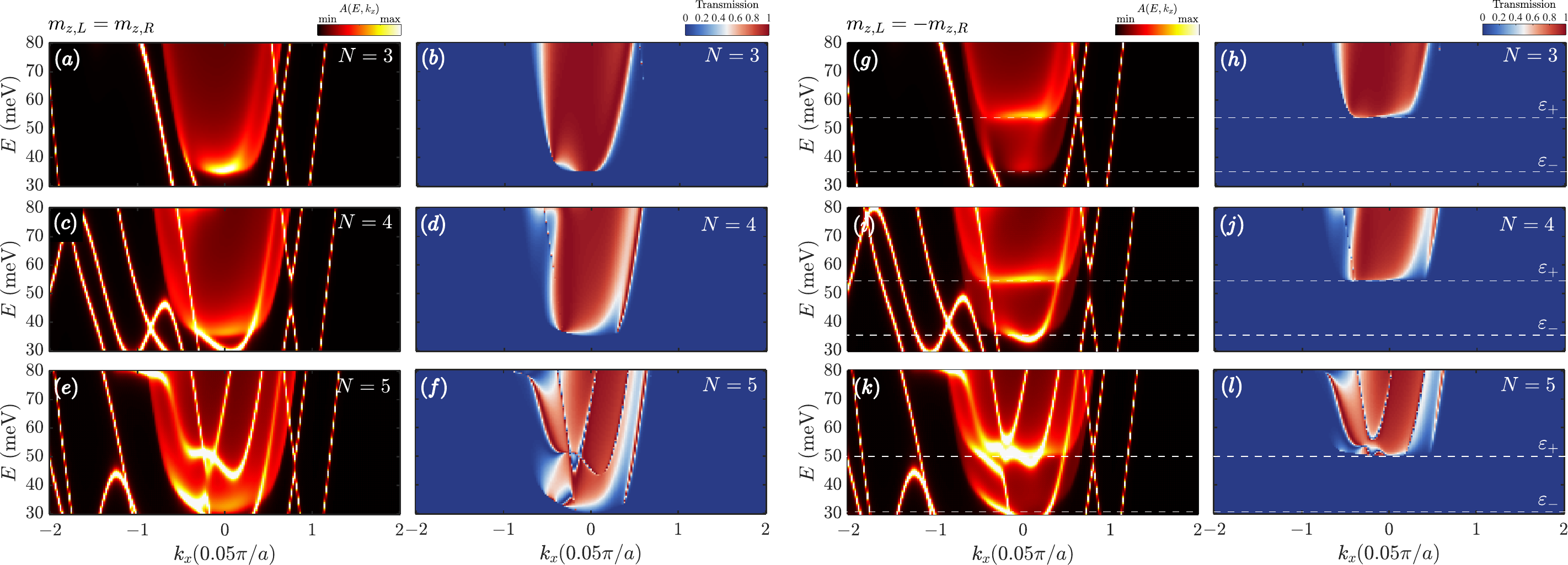}
    \caption{\textbf{Transmission through finite-width domain walls along the zigzag direction}. Spectral function $A(E,k_x)$ and transmission  across zigzag domain walls for $N = 3$ (a,b), $N=4$ (c,d), and $N=5$ (e,f) where the mass does \textit{not} switch sign across the domain walls $m_{z,L} = m_{z,R}$. Spectral function $A(E,k_x)$ and transmission  across zigzag domain walls for $N = 3$ (g,h), $N=4$ (i,j), and $N=5$ (k,l) where the mass does switch sign across the domain walls $m_{z,L} = -m_{z,R}$. Here, $m_z = 10$ meV for all cases, $\Delta = 45,30,20$ meV for $N = 3,4,5$ respectively, and $w = 50a.$ Momenta are measured relative to $\mathbf{K}_+.$}
    \label{fig:zigzag_domain_wall}
\end{figure*}

\begin{figure*}
    \centering
    \includegraphics[width=\linewidth]{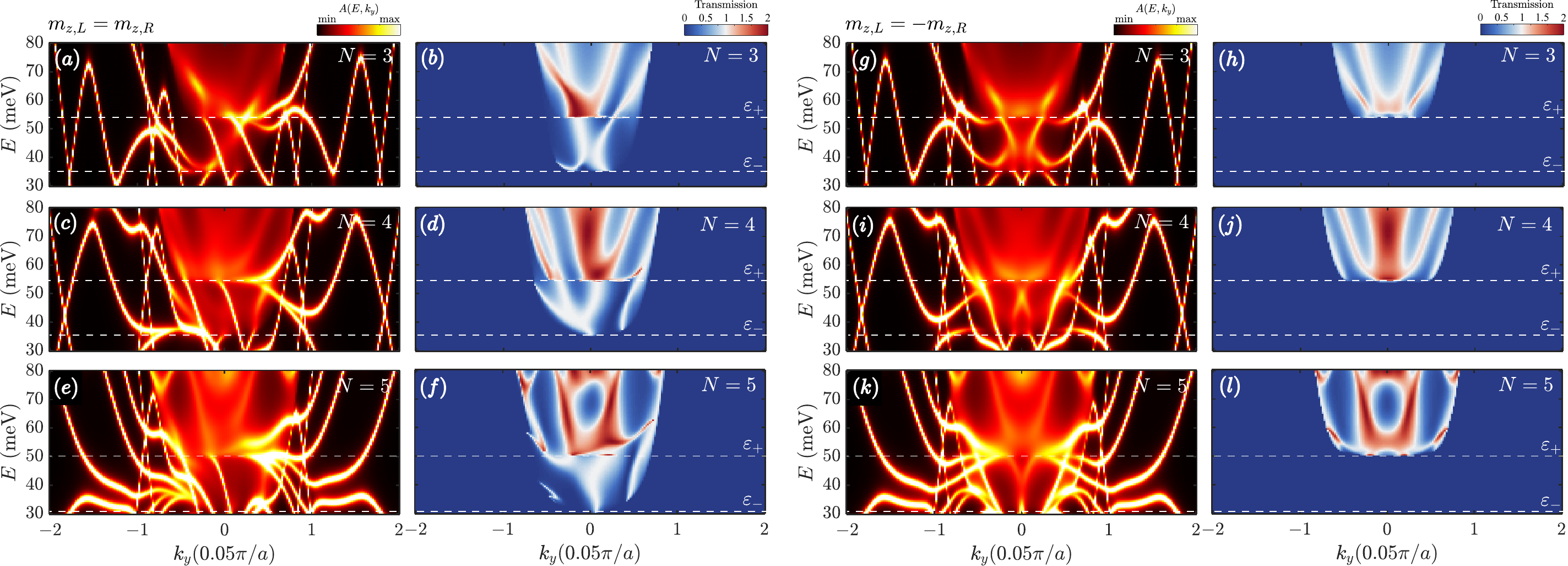}
    \caption{\textbf{Transmission through finite-width domain walls along the armchair direction}. Spectral function $A(E,k_y)$ and transmission  across armchair domain walls for $N = 3$ (a,b), $N=4$ (c,d), and $N=5$ (e,f) where the mass does \textit{not} switch sign across the domain walls $m_{z,L} = m_{z,R}$. Spectral function $A(E,k_y)$ and transmission  across armchair domain walls for $N = 3$ (g,h), $N=4$ (i,j), and $N=5$ (k,l) where the mass does switch sign across the domain walls $m_{z,L} = -m_{z,R}$. Here, $m_z = 10$ meV for all cases, $\Delta = 45,30,20$ meV for $N = 3,4,5$ respectively, and $w = 50a.$}
    \label{fig:armchair_domain_wall}
\end{figure*}

To calculate transmission, we use the Green's function method together with the Caroli-Fisher-Lee formula \cite{caroli1971direct,Fisher1981Relation,Guinea1983Effective,sancho1985highly}. For zigzag domain walls with width $w = 50a \approx 12$ nm, the results for momenta near $\mathbf{K}_+$ are shown in Fig. \ref{fig:zigzag_domain_wall} for $N = 3,4,5$ where $m_z$ has the same sign (a-f) and opposite signs (g-l) across the walls. In Fig. \ref{fig:zigzag_domain_wall}(a,c,e,g,i,k), we show the spectral function of the domain walls, where in addition to the bulk continuum states, we also observe the proliferation of many dispersive wall-localized states induced by the weakening of domain-wall bonds due to tensile strain \cite{zhang2013valley,ju2015topological}. Because of valley polarization $m_z \neq 0,$ we also observe two distinct band edges in the continuum spectra of Fig. \ref{fig:zigzag_domain_wall}(g,i,k) due to a relative shift of the valleys on the two sides of the domain wall. These shifts are absent in Fig. \ref{fig:zigzag_domain_wall}(a,c,e) because, in these cases, the valley polarization is the same on both sides. For $m_{z,L} = m_{z,R},$ we observe from Fig. \ref{fig:zigzag_domain_wall}(b,d,f) that in regions of phase space where there are bulk states, the transmission is not only nonzero but can routinely  reach unity, validating the results from the analytic model. It is worth pointing out that wall-localized states can act as scattering centers and consequently affect transmission in nontrivial ways, as shown most prominently in Fig. \ref{fig:zigzag_domain_wall}(f). For $m_{z,L} = -m_{z,R},$ the bulk states on the two sides of the domain wall are mismatched in energies, as mentioned earlier. Thus, we observe from Fig. \ref{fig:zigzag_domain_wall}(h,j,l) that transmission is only possible for energies larger than the higher-energy conduction band edge $E\geq \varepsilon_+$. In these cases, both sides of the domain walls feature the same valley at the same zone corner, and thus, transmission is not surprising. As before, wall-localized states can act as scatterers that alter the transmission behavior of the bulk states.

In Fig. \ref{fig:armchair_domain_wall}(a,c,e,g,i,k), we show the spectral function of the domain walls along the armchair direction, where in addition to the bulk continuum states, we also observe the proliferation of many dispersive wall-localized states induced by the modification of domain-wall bonds due to shear strain \cite{zhang2013valley,ju2015topological}. We observe two conduction band minima at $\varepsilon_\pm$  because of valley polarization. For $m_{z,L} = m_{z,R},$ the same valleys on the two sides of the domain wall are spectrally aligned but the chiralities are mismatched. We find two regimes: for $\varepsilon_- \leq E < \varepsilon_+,$ the transmission comes entirely  from states at the $\xi=+$ valley, and for $E \geq \varepsilon_+,$ there is transmission from both valleys. This validates again that a mismatch in chirality does \textit{not} dictate blockade behavior. For $m_{z,L} = -m_{z,R},$ opposite valleys on the two sides of the domain wall are now spectrally aligned. In the energy window $\varepsilon_- \leq E<\varepsilon_+$, states on both sides of the domain wall have the same chirality. Despite that, we observe from Fig. \ref{fig:armchair_domain_wall}(h,j,l) that there is practically zero transmission for $E < \varepsilon_+.$

\section{Continuum elasticity modeling of structural domain walls}
\label{sec: Continuum elasticity modeling of quenched domain walls}

This section outlines the balance between elastic and adhesion energies in a graphene multilayer, a competition that governs the equilibrium configuration of structural domain walls, also called here stacking solitons, or solitons for short. We first evaluate the fundamental case of a graphene bilayer, establishing a framework that readily generalizes to multilayer systems. The formalism here builds upon the approach developed in Ref.~\cite{Yankowitz:NM14}.

\subsection{Bilayer case}

In a graphene bilayer, a stacking soliton represents the interface separating AB and BA stacked domains, positioned at $x\to-\infty$ and $x\to\infty$, respectively. The soliton is characterized by an in-plane interlayer displacement field $\mathbf{u}(\mathbf{r}) = \mathbf{u}_{\rm top}(\mathbf{r})-\mathbf{u}_{\rm bot}(\mathbf{r})$, subject to the boundary conditions
\begin{eqnarray}
\mathbf{u}(x\to-\infty)&=&0,\\
\mathbf{u}(x\to\infty)&=&(1+3m)\mathbf{b}_n.
\end{eqnarray}
This shift enforces the asymptotic $AB$ and $BA$ stacking limits, since any interlayer displacement $Mb_n$ with $M\rm{mod}\,3=1$ converts from $AB$ to $BA$ stackings. The integer $m$ classifies solitons of increasing tension, and  $n=0,1,2$ specifies degenerate soliton ``flavors,''. Here $\mathbf{b}_n = (\sin(2\pi n/3), \cos(2\pi n/3), 0)b$ represent the nearest-neighbor bond vectors in the top layer, where $b=|\mathbf{b}_n| = a/\sqrt{3}=0.142$ nm is the carbon-carbon distance. For the initial derivation, we allow only the top layer to deform, keeping the bottom layer rigid.

The objective is to determine the displacement field $\mathbf{u}(\mathbf{r})$ that minimizes the total energy $F = F_0+F_S$. This functional comprises the elastic strain energy of the deformed upper layer,
\begin{eqnarray}
F_0&=&\frac{1}{2}\int d^2r \left[\lambda \left(\sum_{i}u_{ii}\right)^2+2\mu \sum_{ij}u_{ij}u_{ji}\right],\\
u_{ij}&=&\frac{1}{2}\left(\partial_i u_j+\partial_j u_i\right),
\end{eqnarray}
and the van der Waals stacking (or adhesion) energy,
\begin{equation}
F_S=\int d^2r V(\mathbf{u}(\mathbf{r})).
\end{equation}
Currently, $F_0$ depends exclusively on the top layer's deformation. This treatment will be expanded below to accommodate symmetrically distributed strains in bilayers ($\mathbf{u}_{\rm top}=-\mathbf{u}_{\rm bot} = \mathbf{u}/2$) as well as $N$-layer rhombohedral systems.

The adhesion term $F_S$ originates from the area-normalized energy cost of specific \emph{uniform} stacking configurations, defined by $V(\mathbf{u})$. The $AB$ and $BA$ registries correspond to global minima, such that $V(\mathbf{0})=V(\mathbf{b}_n)=0$. Any deviation from these registries incurs an energy penalty. The inclusion of $F_S$ prevents the soliton from expanding indefinitely, which would otherwise happen if the system solely minimized the elastic penalty $F_0$.

\begin{figure}[t] 
   \centering
   \includegraphics[width=0.49\textwidth]{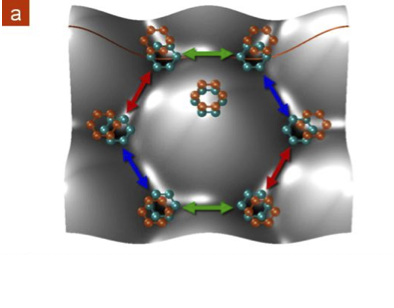}
   \includegraphics[width=0.49\textwidth]{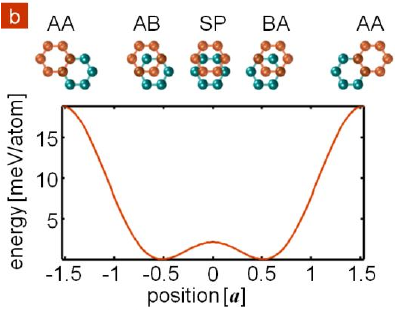}
   \caption{Adapted from Ref.~\cite{Alden:PNAS13}: (a) Area-normalized uniform stacking energy $V(\mathbf{u})$ as a function of interlayer displacement $\bm{u}=(u_x,u_y)$. The interlayer registry (stacking alignment) at each high-symmetry point are shown. (b) Cross-section of the potential landscape along the $AA-AB-BA-AA$ registry trajectory.}
   \label{fig:V}
\end{figure}

Following Ref.~\cite{Alden:PNAS13} (see Fig.~\ref{fig:V}), we model $V(\mathbf{u})$ assuming it imposes a steep energy penalty for displacements outside the three specific directions $\mathbf{b}_n$. The potential evaluated along these particular channels, illustrated in Fig.~\ref{fig:V}(b), reduces to:
$$
V(z\mathbf{b}_n)=\mathcal{V}(z).
$$
Because alternative displacement paths are energetically restricted, the soliton displacement field is effectively constrained to:
$$
\mathbf{u}(\mathbf{r}) = \mathbf{b}_n f(\mathbf{r})
$$

The scalar function $f(\mathbf{r})$ encapsulates the spatial profile of the domain wall, satisfying the asymptotic constraints:
\begin{eqnarray}
f(x\to-\infty)&=&0,\\
f(x\to\infty)&=&1+3m.
\end{eqnarray}
If $m=0$ we have the lowest-energy soliton, that corresponds to an $AB/$SP$/BA$ domain wall, here dubbed an SP-soliton (SP stands for ``saddle-point'' stacking, see Fig. \ref{fig:V}). The $m=-1$ soliton, in contrast, corresponds to an $AA$-soliton, where stacking evolves in an $AB/AA/BA$ sequence. Within the collinear ansatz it is the next stationary solution in order of energy.
We can define a rescaled profile function 
$$
g(\bm{r})=\frac{2}{1+3m}f(\bm{r})-1
$$ that has $m$-independent boundary conditions $g(x\to-\infty)=-1$ and $g(x\to\infty)=1$ for any $m$. This is the parametrization used in SM Sec. \ref{sec: Atomistic Tight-Binding Model}.

The soliton ansatz is simplified by modeling it as a straight ridge rotated by an angle $\theta$ relative to $\mathbf{b}_n$. This allows us to reduce the problem from 2D to 1D and write
$$
f(\mathbf{r})=f(\mathbf{r}\cdot\mathbf{\hat m}_\theta),
$$
where $\mathbf{\hat m}_\theta$ represents the unit vector perpendicular to the domain wall, and $f$ is now a function of the distance $\mathbf{r}\cdot\mathbf{\hat m}_\theta$ to the soliton center only. For analytical convenience, we choose our axes so that $\mathbf{b}_n=\hat{y}a/\sqrt{3}$. The normal vector is therefore \[\mathbf{\hat m}_\theta=(\cos\theta, \sin\theta).\]

Evaluating the strain tensor for this geometry yields:
$$
u_{ij}(\mathbf{r})=b\left(\begin{array}{cc}
0&\frac{1}{2}\cos\theta\\
\frac{1}{2}\cos\theta&\sin\theta
\end{array}
\right)f'(\mathbf{r}\cdot\mathbf{\hat m}_\theta).
$$

Integrating this provides the elastic energy functional:
$$
F_0=\frac{b^2}{2}\left(\mu+B\sin^2\theta\right)\int d^2 r\left[f'(\mathbf{r}\cdot\mathbf{\hat m}_\theta)\right]^2.
$$
Here, $B=\lambda+\mu\approx 12.6~\mathrm{eV/\AA^2}$ denotes the 2D bulk modulus of a single graphene layer, and $\mu\approx 9~\mathrm{eV/\AA^2}$ corresponds to its shear modulus. Considering $F_0$ with a fixed profile $f(\mathbf{r}\cdot\mathbf{\hat m}_\theta)$ and varying $\theta$ we see that the elastic energy reaches a minimum at $\theta=0$ (the ``shear'' soliton) and a maximum at $\theta=\pi/2$ (the ``tensile'' soliton).

\begin{figure}[] 
   \centering
   \includegraphics[width=0.5\columnwidth]{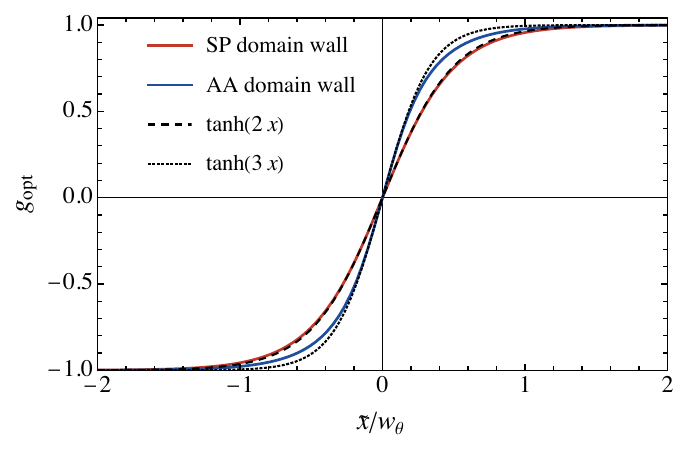}
   \caption{Normalized spatial profile $g_{\rm{opt}}(x)$ of the equilibrium SP and $AA$ soliton solutions, as a function of normalized distance from the soliton center $x = \tilde{x}/w_\theta$. The characteristic length $w_\theta$ is calculated in Eq.~(\ref{W}). The dashed/dotted lines are analytic approximations.}
   \label{fig:profile}
\end{figure}

By introducing a local coordinate system oriented across ($\tilde{x}=\mathbf{r}\cdot\mathbf{\hat m}_\theta$) and parallel to the domain wall ($\tilde{y}=\mathbf{r}\cdot[\mathbf{\hat{z}}\times\mathbf{\hat m}_\theta]$), we isolate the shape profile $f(\tilde{x})$ required to minimize $F=F_0+F_S$. Because the integrand is invariant with respect to $\tilde{y}$, integration along that axis simply factors out the total soliton length $L$. The reduced 1D energy functional becomes:
\begin{equation}
F=L\int d\tilde x\left\{\mathcal{V}[f(\tilde x)]+ \frac{b^2}{2}\left(\mu+B\sin^2\theta\right)\left[f'(\tilde x)\right]^2\right\}
\label{Fx}.
\end{equation}

We adopt the parameterized form $\mathcal{V}(z)=\mathcal{V}_0\mathcal{\tilde V}(z)$, wherein
$$
\mathcal{\tilde V}(z)\approx \left[1-2\cos\left(\frac{\pi}{3} (2z-1)\right)\right]^2,
$$
with the maximum energy barrier extracted from Fig.~\ref{fig:V}(b) as $\mathcal{V}_0\approx 2 \mathrm{ meV}/\textrm{atom}=1.52 \mathrm{meV/\AA^2}$.

Equation (\ref{Fx}) can be rewritten in terms of dimensionless variables as
\begin{equation}
F=\mathcal{V}_0 L w_\theta \int_{-\infty}^{\infty} \frac{d\tilde x}{w}\left\{\mathcal{\tilde V}[f(\tilde x)]+ \left[wf'(\tilde x)\right]^2\right\}
\label{Fx2}
\end{equation}
by defining the characteristic scale
\begin{equation}
w_\theta=\sqrt{\frac{\mu+B\sin^2\theta}{2\mathcal{V}_0}}b,
\label{W}
\end{equation}
which quantifies the typical domain wall width. This length is $w_{0}=7.7$ nm for the shear configuration and $w_{\pi/2}=12$ nm for the tensile case (both assuming a pinned substrate, see below for the suspended multilayer generalization). The dimensionless scale $2\mathcal{V}_0/(\mu+B\sin^2\theta)\sim 10^{-4}$ is found to control a wide variety of structural scales in graphene multilayers, from the relaxation of moiré patterns in twisted bilayers \cite{nam_lattice_2017,guinea_continuum_2019,koshino_moire_2019,ochoa_moire-pattern_2019,ezzi_analytical_2024,de_beule_theory_2026} to the aspect ratio of graphene bubbles \cite{khestanova_universal_2016}.

The equilibrium profiles $f$ that minimize Eq.~(\ref{Fx2}) obey the spatial scaling relation $f(\tilde x)=f_\mathrm{opt}(\tilde x/w_\theta)$. Altering physical parameters like $\theta$ or $\mathcal{V}_0$ simply modulates the effective width $w_\theta$ via this universal function $f_\mathrm{opt}(x)$ (where $x=\tilde x/w_\theta$ is the normalized distance to the soliton) without changing the fundamental soliton profile. We compute numerically the optimal solution $f_\mathrm{opt}(x)$. Its rescaled version $g_\mathrm{opt}(x)=2f_\mathrm{opt}(x)/(1+3m)-1$ is plotted in Fig.~\ref{fig:profile} for $m=0$ (SP-soliton) and $m=-1$ ($AA$-soliton). We find that the latter is slightly narrower than the former, since it pays a higher adhesion energy penalty at its center. They are well approximated by a $\tanh(2x)$ and $\tanh(3x)$ soliton profile, respectively, as shown by the dashed and dotted lines. If we quantitatively define the full-width-at-half-maximum (FWHM) with the condition $g(\tilde x/w_\theta)^2=1/2$, we find that $w_\theta$ is indeed quite close to the typical FWHM for the SP soliton (13\% larger to be precise), and around 1.7 times that for the narrower $AA$-soliton.


Substituting the relaxed profile back into the functional yields the optimal energy:
\begin{equation}
F_\mathrm{opt}=\mathcal{V}_0 L w_\theta \int_{-\infty}^{\infty} dz\left\{\mathcal{\tilde V}[f_\mathrm{opt}(z)]+ \left[f_\mathrm{opt}'(z)\right]^2\right\}.
\end{equation}

\subsection{Multilayer case}

We now expand this derivation to an $N$-layer rhombohedral system, with a domain wall between a left-side $ABCABCA\dots$ sequence ($x\to-\infty$) and a right-side $BACBACB\dots$ sequence ($x\to\infty$). The nature of rhombohedral crystals implies that every individual interface (out of $N-1$ total) transitions from AB stacking to BA stacking across the multilayer soliton. There are multiple ways to achieve this, in particular any relative interlayer shift between layers $\ell$ and $\ell+1$ of the form $(1+3m_\ell) \bm{b}_{n_\ell}$ (with $n_\ell=0,1,2$ and integer $m_\ell$. These types of solitons are specified by pairs $(n_\ell,m_\ell)$ for each pair of adjacent layers and cannot in general be locally deformed one into the other, so they should be stable for their respective optimal deformation profile. The actual soliton configuration that  will occur in real samples depends on the kinetics of domain wall formation when fabricating the sample. Instead of attempting to model such process, we explore the equilibrium energy and profile for the simplest kind of domain wall corresponding to a layer-independent integer $m_\ell=m$ and any fixed $n_\ell=n=0,1,2$. The relative displacement between any adjacent pair of layers then matches the bilayer case: $\mathbf{u}_{\ell+1}(\mathbf{r})-\mathbf{u}_\ell(\mathbf{r})=f(\mathbf{r})\mathbf{b}_n$, evaluated for $\ell=1,\dots,N-1$. 

\subsubsection{Suspended}

If the multilayer is suspended, the net shift vanishes everywhere, $\sum_\ell\mathbf{u}_\ell(\bm{r})=0$ (the generalized analogue to the symmetric $\mathbf{u}_{\rm top}=-\mathbf{u}_{\rm bot}$ case in bilayers). Then the absolute displacement field for the $\ell$-th lattice plane becomes:
$$\mathbf{u}_\ell(\mathbf{r})=\left(\ell - \frac{N+1}{2}\right)f(\mathbf{r})\mathbf{b}_n.$$

The aggregate elastic penalty $F_0^{(N)}$ is computed by summing the individual in-plane deformation energies for all $N$ crystalline sheets. Because intralayer strain scales linearly with local displacement, the total multilayer elastic energy is:
\begin{eqnarray*}
    F_0^{(N)}&=&\sum_{\ell=1}^N\left(\ell-\frac{N+1}{2}\right)^2F_0=\frac{(N^2-1)N}{12}F_0,
\end{eqnarray*}
where $F_0$ refers back to the bilayer elastic energy with the bottom layer pinned to the substrate \footnote{One important caveat in the above generalization to $N>2$ layers is that we have assumed that the stacking energy density is a sum of adjacent layers only. Hence, the stacking energy difference between $ABA$ and $ABC$ is neglected here. Improving over such approximation leads to a much more complicated phenomenology that could make domain walls between rhombohedral domains reconstruct in non-trivial ways.}. Concurrently, the overall stacking energy scales linearly with the number $N-1$ of layer interfaces:
$$F_S^{(N)}=(N-1)\int d^2 r\,\mathcal{V}[f(\mathbf{r})].$$
Combining these terms constructs the total thermodynamic potential for the $N$-layer defect:
\begin{eqnarray*}
F^{(N)}&=&(N-1)L\int d\tilde{x}\\
&&\left\{\mathcal{V}[f(\tilde{x})]+\frac{(N+1)N}{12}\frac{b^2}{2}\left(\mu+B\sin^2\theta\right)[f'(\tilde{x})]^2\right\}.
\end{eqnarray*}

This expression maps directly onto the dimensionless energy functional Eq.~(\ref{Fx2}) with the effective width rescaling as
\begin{equation}
    w^{(N)}_\theta=\sqrt{\frac{(N+1)N}{12}}w_\theta,
\end{equation}
where $w_\theta$ is the rigid-substrate width defined in Eq.~(\ref{W}). For a fully relaxed bilayer ($N=2$), this yields $w^{(2)}_\theta=w_\theta/\sqrt{2}$. For $N=3$ we have $w^{(3)}_\theta = w_\theta$, i.e. the soliton width in a suspended trilayer is the same as in a bilayer with a pinned substrate, 7.7 nm (shear) to 12 nm (tensile). As $N$ increases, the width of the multilayer domain wall considered here broadens monotonically.

\subsubsection{Pinned to a substrate}
If the multilayer is deposited on a substrate that pins the bottom layer $\ell=1$, so that its deformation is zero, the soliton deformation ansatz becomes
$$\mathbf{u}_\ell(\mathbf{r})=\left(\ell - 1\right)f(\mathbf{r})\mathbf{b}_n.$$
The $F_S^{(N)}$ is exactly the same as above, but the elastic energy $F_0^{(N)}$ is now
\begin{eqnarray*}
    F_0^{(N)}&=&\sum_{\ell=1}^N\left(\ell-1\right)^2F_0=\frac{(N-1)(2N-1)N}{6}F_0,
\end{eqnarray*}
so that 
\begin{equation}
    w^{(N)}_\theta=\sqrt{\frac{(2N-1)N}{6}}w_\theta,
\end{equation}
If $N=2$ we recover the result from the previous section $w^{(2)}_\theta=w_\theta$, while for $N=3$ we have soliton widths that range from 12.2 nm (shear) to 19 nm (tensile). The quantitative FWHM for the pinned trilayer SP and $AA$ solitons yields 10.7 nm (SP-shear) to 16.7 nm (SP-tensile) and 7.1 nm ($AA$-shear) to 11.1 nm ($AA$-tensile).

\end{document}